\documentclass[fleqn,usenatbib]{mnras}

\usepackage{newtxtext,newtxmath}
\usepackage{acronym}
\usepackage{bm}

\usepackage[T1]{fontenc}

\DeclareRobustCommand{\VAN}[3]{#2}
\let\VANthebibliography\thebibliography
\def\thebibliography{\DeclareRobustCommand{\VAN}[3]{##3}\VANthebibliography}

\usepackage{graphicx}	
\usepackage{amsmath}	
\usepackage{multirow}
\usepackage{comment}

\title[Varaha]{Varaha: A promising sampler for obtaining gravitational wave posteriors.}

\author[]{
Vaibhav Tiwari,$^{1}$\thanks{E-mail:v.tiwari@bham.ac.uk}
\\
$^{1}$Institute of Gravitational Wave Astronomy, School of Physics and Astronomy, University of Birmingham, Edgbaston, Birmingham B15 2TT, UK
}

\date{Accepted XXX. Received YYY; in original form ZZZ}

\pubyear{2023}

\begin{document}

\newboolean{comments}   
\setboolean{comments}{true}
\ifthenelse{\boolean{comments}}{
\newcommand{\steve}[1]{\textcolor{magenta}{[Steve: #1]}}
\newcommand{\charlie}[1]{\textcolor{blue}{[Charlie: #1]}}
\newcommand{\vaibhav}[1]{\textcolor{cyan}{[Vaibhav: #1]}}
}{
\newcommand{\steve}[1]{{}}
\newcommand{\charlie}[1]{{}}
\newcommand{\duncan}[1]{{}}
\newcommand{\vaibhav}[1]{{}}
}

\newcommand{\D}{\mathrm{d}}

\label{firstpage}
\pagerange{\pageref{firstpage}--\pageref{lastpage}}
\maketitle

\begin{abstract}
Nested sampling is often used in Bayesian statistics problems in astronomy. It operates with a set of live points, iteratively replacing the point with the lowest likelihood with a new point of higher likelihood. Each iteration reduces the enclosed volume by a known factor. The estimated sampling density and the likelihood values of both new and old live points quantify the enclosed probability mass. Although robust, nested sampling often discards a majority of the sampled points~($\sim 99.9\%$) at which likelihood was calculated. Here, we present an efficient method to explicitly calculate the sampling density for small dimensional problems~(ten or less), thereby removing the need to discard samples. The points' sampling density and likelihood values constitute the posterior distribution. We build on the existing version of the sampler Varaha and present an alternate version that is significantly more efficient for expensive likelihoods. These samplers specifically focus on obtaining compact binary parameters from their gravitational wave signals. They provide a viable alternative to nested sampling when the full fifteen-dimensional space is sampled separately for observer-dependent parameters and parameters intrinsic to the binary.
\end{abstract}


\acrodef{GW}[GW]{gravitational wave}
\acrodef{PE}{Parameter Estimation}
\acrodef{MC}{Monte Carlo}
\acrodef{MCMC}{Markov Chain Monte Carlo}
\acrodef{PDF}{Probability Density Function}
\acrodef{NS}{Neutron Star}
\acrodef{BH}{Black Hole}
\acrodef{BBH}[BBH]{Binary Black Hole}
\acrodef{BNS}[BNS]{Binary Neutron Star}
\acrodef{CBC}{Compact Binary Coalesence}
\acrodef{EDF}{empirical distribution function}
\acrodef{CDF}{cumulative distribution function}
\acrodef{SNR}{Signal to Noise Ratio}



\section{Introduction}

Parameter estimation of compact binaries from their \ac{GW} signal is essential for inferring the compact binary population, conducting tests of general relativity, or performing cosmological studies~(e.g. see \cite{2023ApJ...955..107F, 2023ApJ...957...37R, 2023ApJ...958...13A, 2023arXiv231200993H, 2024ApJ...960...65S, 2024arXiv240207075G, 2024A&A...684A.204R, 2024PhRvX..14b1005C, 2024ApJ...966L..16P, 2024arXiv240402522M, 2024PhRvD.109f4056L, 2023arXiv230408393T, 2024MNRAS.527..298T} for some recent works). 

Compact binary parameters are often estimated using Monte Carlo methods, where the likelihood is calculated for hundreds of millions of data points drawn from the parameter space. Each likelihood calculation can take a few tens to a few thousand milliseconds~\citep{2021PhRvD.103j4056P}. In addition, multiple copies of each analysis are run to collect a sufficient number of posterior samples, and each signal is analysed several times using different waveforms and sometimes with various choices on the prior distribution~\citep{2021arXiv210801045T, o3b_cat}. The overall computational load, although enormous, has been manageable. 

However, with ground-based \ac{GW} detectors becoming increasingly sensitive, the number of observed \ac{GW} signals is expected to be in the hundreds per year~\citep{2020LRR....23....3A}. The total computational cost for estimating the parameters is expected to rise proportionally. Additional computational load is expected due to increased sensitivity. Improvement at lower frequencies will require the generation of longer-duration gravitational waveforms to estimate the parameters. Increased sensitivity will also enable the measurement of new parameters, e.g. eccentricity, which will likely require increased computation. Often, analyses need to be micromanaged for various reasons, and with each analysis finishing over multiple days, the increased detection rate may pose a logistical challenge. 

Significant efforts have been aimed at addressing these challenges. These efforts have focused on making approximate but fast estimations of parameters~\citep{2023PhRvD.108h2006F}, speeding up likelihood calculation, constituting a major fraction of the computational load~(e.g. see \cite{2020PhRvD.102j4020M, 2021PhRvD.104d4062M, 2023PhRvD.108f4055P, 2023PhRvD.108l3040M, 2023arXiv230812140N, 2023PhRvD.108l3025M, 2024PhRvD.109b4053P, 2024arXiv240402435R} for some relevant works.), and improving the sampling efficiency of the analysis using machine learning~\citep{2021PhRvD.103j3006W, 2021PhRvL.127x1103D, 2022NatPh..18..112G, 2023ApJ...958..129W}.

 It is also possible to break the sampling problem into the observer dependent~(extrinsic) parameters and observer independent~(intrinsic) parameters~\citep{2015PhRvD..92b3002P, 2018arXiv180510457L, 2023PhRvD.108b3001T, 2023ApJ...958..129W}. The likelihood of arbitrary extrinsic parameters can be obtained by applying amplitude/phase shifts to the waveform for fixed intrinsic parameters. The likelihood is first marginalised to obtain the posterior over extrinsic parameters for fixed intrinsic parameters. Following this, the intrinsic parameters are sampled using the marginalised likelihoods. The analysis incurs an additional cost for obtaining marginalised likelihoods but requires a significantly smaller number of waveform generations as the sampling over the intrinsic parameters is performed on dimensionally small parameter space.

All parameter estimation analyses benefit from improvements in sampling schemes. We build upon our previous presentation~\citep{2023PhRvD.108b3001T} and report an alternate version of Varaha; a version highly efficient in small dimensions for expensive likelihoods.

\section{Method}
\label{sec:method}
Nested sampling is a robust tool for estimating gravitational wave posterior. Nested sampling converges to the peak of the likelihood distribution by iteratively drawing a series of contours of increasing likelihood. Initially, it draws a set of \emph{live points}, $n_\mathrm{live}$. At each cycle, it stores the live point with the lowest likelihood and replaces it with a new point drawn randomly from the prior. The new point must have a likelihood value greater than the smallest likelihood value of the live points. This scheme results in volume shrinkage by an approximate factor of $\sim e^{-1/n_\mathrm{live}}$. The new live points advance to a region with a higher likelihood. The stored points, along with their weight -- the product of the average likelihood value and the volume enclosed between successive contours -- constitute the \emph{samples} drawn from the posterior distribution. Nested sampling suffers from two drawbacks: i) it has to cross-reference the likelihood values at each iteration, making likelihood calculation impractical to parallelise over many CPUs, and ii) a live point is replaced only after a point of higher likelihood has been obtained. This results in discarding a large number of points where likelihood is calculated. When estimating compact binary parameters from gravitational waves, most of the points where likelihood was calculated are often discarded~($\sim 99.9\%$). 

Parallel Bilby improves the process parallelisation by independently advancing live points on multiple CPU nodes. On obtaining points higher in likelihood, all the points are collected, and new live points are defined~\citep{2020MNRAS.498.4492S}. Thus, Parallel Bilby can shrink the volume by a factor greater than $\sim e^{-1/n_\mathrm{live}}$ at each iteration. However, Parallel Bilby is better suited for expensive likelihoods as the overhead cost of invoking parallelisation at each iteration may become high. The likelihood calculations also increase as not all the points accepted on different CPU nodes eventually get included as live points.

For general problems, nested sampling provides an astute methodology to properly estimate the sampling density at each iteration. A possible way to improve the sampling efficiency would be to keep all the points and to explicitly calculate their sampling density. This approach is expected to become cumbersome with an increase in the number of parameters. However, the number of parameters for compact binaries is often not large. For example, there are 15 parameters for a precessing binary~(8 and 7 if the sampling is broken into extrinsic and intrinsic parameters)~\citep{2019ApJS..241...27A}. The inclusion of calibration errors significantly increases the number of parameters~\citep{first_monday_pe}, but as none of the additional parameters are expected to be correlated with the binary parameters, these can be treated separately~\citep{2020PhRvD.102l2004P}.

To explicitly calculate the sampling density, we build a recipe based on a previous work~\citep{2023PhRvD.108b3001T}. We review the previous work as follows. We begin by sprinkling the full parameter space with $N$ points. On selecting $n$ points with the largest likelihood values, we carve a volume enclosed by the minimum likelihood value among all the selected points, $\mathcal{L}_*$. This volume, $V(\mathcal{L}_*)$ can be estimated using \ac{MC} integration, 
\begin{equation}
    V(\mathcal{L}_*) \approx \bar{V}(\mathcal{L}_*) = V_0\frac{n}{N}, \quad \delta \bar{V} = \frac{\bar{V}}{\sqrt{n}},
\end{equation}
where $V_0$ is the full volume of the parameter space and $\delta \bar{V}(\mathcal{L}_*)$ is the error on the estimated volume, $\bar{V}(\mathcal{L}_*)$. The structure of this volume is not known. But, on binning the full parameter space with bin size equal to $\delta \bar{V}(\mathcal{L}_*)$ and selecting the bins that include at least one of the selected points, we reconstruct a volume~(referred as \emph{live volume} from now on). Live volume is expected to enclose most of the -- if not all -- \emph{true} volume. We are focused on Gaussian likelihood~\citep{first_monday_pe} when estimating \ac{GW} posterior. For this likelihood, the distribution smoothly decays from a peak value. The error in the posterior probability is significantly smaller than in the volume -- which is expected to be incurred away from the bulk probability in the lower-likelihood regions.

\begin{center}
\begin{figure*}
\includegraphics[width=\textwidth]{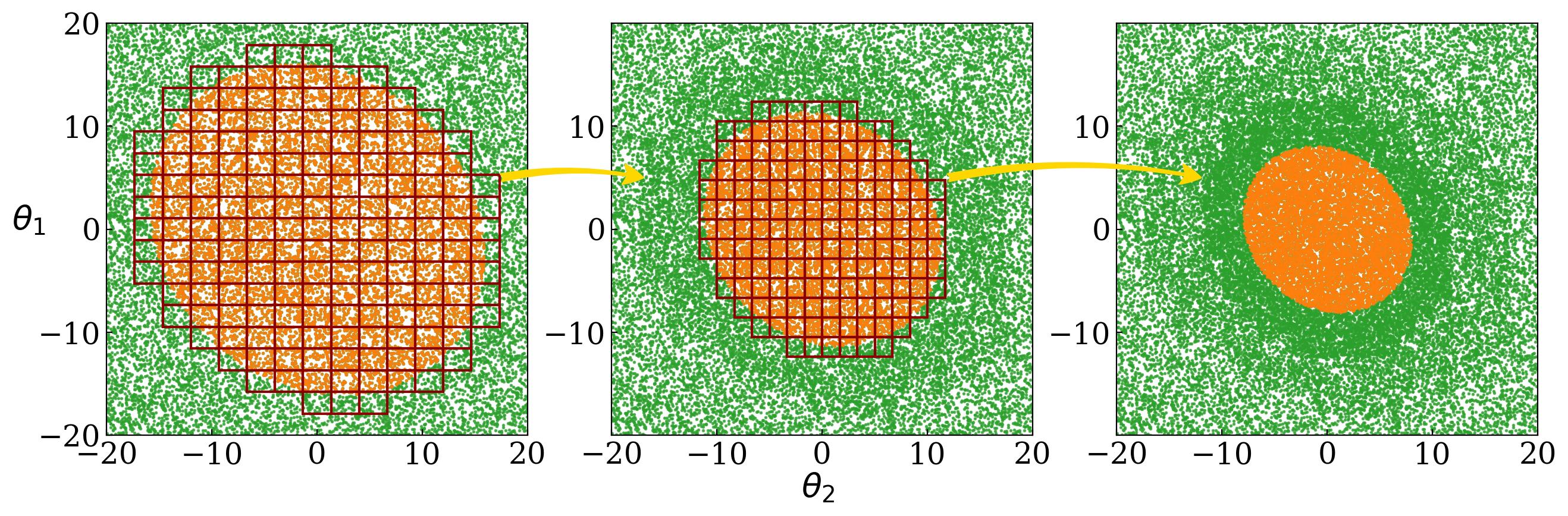}
\vspace{-1em}
\caption{Sprinkling $N$ points~(green point) in a volume and selecting $n$ points~(orange points) with largest likelihood values carves a volume that is approximately $V(\mathcal{L}_*) \sim n/N$. The error associated with this volume is $\delta V(\mathcal{L}_*) \sim 1/\sqrt{n}$. Binning the parameter space with bin size equal to $\delta V$ and selecting bins that enclose at least one of the selected points~(red grid) ensures the maximum \emph{true} volume not included by the selected bins is $\delta V$. For \emph{Gaussian likelihoods}, error on volume always implies a significantly small error in probability mass enclosed by the likelihood threshold $\mathcal{L}_*$. Thus, only sampling from the bins at the next cycle ensures the parameter space enclosed by $\mathcal{L}_*$ has been sufficiently sampled. The analysis keeps evolving until a stopping criterion is met.}
\label{fig:evolution}
\end{figure*}
\end{center}

Thus, just sampling from the bins will, in practice, increase the number of data points with $\mathcal{L} > \mathcal{L}_*$ while ensuring that all the parameter space enclosed by the likelihood threshold $\mathcal{L}_*$ has been sampled. From the second cycle onwards, the procedure is repeated while only considering points with $\mathcal{L} > \mathcal{L}_*$ and discarding the rest. This is outlined pictorially in Fig.~\ref{fig:evolution}. The likelihood threshold, $\mathcal{L}_*$, is allowed to evolve until the probability mass discarded by the analysis reaches a pre-chosen threshold~\citep{2023PhRvD.108b3001T}. The likelihood threshold is not decreased further, and the enclosed volume is repeatedly sampled until a desired number of posterior samples are collected. As the bins cover a volume that encloses almost all probability mass, keeping track of sampling density is unnecessary.

However, because the reconstructed live volume is much larger than the true live volume and as a majority of the samples are drawn from the lower likelihood region of the live volume, the sampling efficiency, especially in large dimensions, is low. In the presented analysis, we evolve the likelihood threshold to the peak of the likelihood distribution. This results in increased sampling efficiency as the bins now densely sample the high-likelihood regions of the likelihood distribution. However, we now need to explicitly calculate the sampling density of the sampled points. With $N^i_j$ samples drawn from the $j^{th}$ bin, in the $i^{th}$ iteration/cycle, the sampling density for the $k^{th}$ sample is
\begin{equation}
    \rho_k = \sum_{i} \frac{N^i_j}{\delta \bar{V}^i}\,\Theta\left(j^i, k\right),
    \label{eq:injupd1}
\end{equation}
where $\delta \bar{V}^i$ is the estimated error in the \ac{MC} volume at the $i^{th}$ iteration/cycle. $\Theta\left(j^i, k\right)$ is one when the $k^{th}$ sample overlaps with the $j^{th}$ bin in the $i^{th}$ cycle. Thus, Eq.~\ref{eq:injupd1} obtains the sampling density for a point by adding contributions from overlapping bins in all the cycles. For samples drawn at the current cycle, Eq.~\ref{eq:injupd1} needs to be calculated fully; however, for the samples drawn in previous cycles, the sampling density needs to be updated by the contributions from bins in the current cycle. That is for the current cycle $i$, any 
$k^{th}$ sample drawn in a previous cycle will have an increase in its sampling density by,
\begin{equation}
    \frac{N^i_j}{\,\delta \bar{V}^i}\,\Theta\left(j^i, k\right).
\end{equation}

The samples don't carry equal weights because of different sampling densities. Rather, their weight is given by the inverse of the sampling density,
\begin{equation}
    w_k = \frac{1}{\rho_k}.
\end{equation}
The index $k$ covers all the collected samples. Instead of choosing $n$ points with the largest likelihood values, the analysis selects the next likelihood threshold, $\mathcal{L}_*^{i+1}$, that shrinks the volume by a factor $f$. This new threshold will satisfy,
\begin{equation}
    f = \frac{\sum_k w_k(\mathcal{L}_k > \mathcal{L}_*^{i+1})}{\sum_k w_k(\mathcal{L}_k > \mathcal{L}_*^i)}, \quad \bar{V}^{i+1} = f\, \bar{V}^i, \quad 0 < f < 1,
    \label{eq:vol_next_lkl}
\end{equation}
where the inequality in the brackets puts a condition on the sample to qualify for addition.
The fractional error on the volume $\bar{V}^{i+1}$ is
\begin{equation}
    \frac{\delta \bar{V}^{i+1}}{\bar{V}^{i+1}} = f \sqrt{\frac{\mathrm{var}(A)}{A^2} + \frac{\mathrm{var}(B)}{B^2} - 2\frac{\mathrm{cov}(A, B)}{AB}},
    \label{eq:vol_next_lkl_err}
\end{equation}
where $A$ and $B$ are numerator and denominator in the Eq.~\ref{eq:vol_next_lkl}. Var and Cov stand for variance and covariance. The error is small for large $f$. For small values of $f$, only the first term in Eq.~\ref{eq:vol_next_lkl_err} contributes and the error reduces to,
\begin{equation}
    \delta \bar{V}^{i+1} = \frac{\bar{V}^{i+1}}{\sqrt{n^{i+1}_{\mathrm{eff}}}}
    \label{eq:cons_err}
\end{equation}
where $n^{i+1}_{\mathrm{eff}}$ is the effective prior-sample size enclosed by the likelihood threshold, $\mathcal{L}_*^{i+1}$~\citep{ESS, 2018CQGra..35n5009T},
\begin{equation}
    n^{i+1}_{\mathrm{eff}} = \frac{\left(\sum_k w_k(\mathcal{L}_k > \mathcal{L}_*^{i+1})\right)^2}{\sum_k \left(w_k(\mathcal{L}_k > \mathcal{L}_*^{i+1})\right)^2}.
\end{equation}
The value of the likelihood threshold increases with each cycle. The volume enclosed by the likelihood threshold, $\mathcal{L}_*^i$ gradually decreases as the cycles proceed~(for the full volume $V_0$, the corresponding likelihood threshold is -$\infty$),
\begin{equation}
    \bar{V}(\mathcal{L}^l_*) = V_0 f^l = V_0\prod_{i=0}^{l-1}\frac{\sum_k w_k(\mathcal{L}_k > \mathcal{L}_*^{i+1})}{\sum_k w_k(\mathcal{L}_k > \mathcal{L}_*^i)} = V_0\frac{\sum_k w_k(\mathcal{L}_k > \mathcal{L}_*^l)}{\sum_k w_k}.
    \label{eq:nested_vol}
\end{equation}
In Eq.\ref{eq:vol_next_lkl_err}, we have ignored the error in $\bar{V}^i$. To accommodate this error, we shrink the volume by a fraction close to one but create bins of the same size as the conservative estimate of the volume error. This is given in Eq.~\ref{eq:cons_err}. Eq.\ref{eq:vol_next_lkl} suggests that accurate measurement of $\mathcal{L}_*^{i+1}$ depends on $\sum_k w_k(\mathcal{L}_k > \mathcal{L}_*^i)$ and the error $\delta \bar{V}^{i+1}$ depends on $n^i_{\mathrm{eff}}$\footnote{Please note, $\sum_k w_k(\mathcal{L}_k > \mathcal{L}_*^i) / \max(w_k(\mathcal{L}_k > \mathcal{L}_*^i))$ is reduced prior-sample size, which informs the approximate number of equal weight samples one can obtain by performing rejection sampling on the reduced sampling weights, $w_k(\mathcal{L}_k > \mathcal{L}_*^i) / \max(w_k(\mathcal{L}_k > \mathcal{L}_*^i))$. Moreover, $n^{i+1}_{\mathrm{eff}} \approx fn^i_{\mathrm{eff}}$.}. 

The effective prior-sample size indicates how densely the space has been sampled. The reduced prior sample size indicates how disparate the sampling densities are in the sampled space. If sampling is rigorous, the two are expected to be proportional. We define a volume enclosed by a likelihood threshold sufficiently sampled by requiring them to be greater than some pre-chosen desired numbers. As we can only control $\delta \bar{V}^i$ in the analysis, the bins may not adequately sample the parameter space if there are features in the distribution that are smaller than this error. Thus, care must be taken when making the choices. For the examples presented in this article, we have chosen an effective prior-sample size greater than 10,000, which ensures a conservative estimate of the error on volume to be approximately 1\%. But, as large errors are related to small values of $f$, we choose $f=0.95$ to ensure that the error is much smaller. We also require a reduced sample size greater than 2,000. The reduced sample size indicates the number of independent samples enclosed by the likelihood threshold. This resembles nested sampling, where the number of independent samples enclosed by the likelihood threshold is the number of live points. Thus, our choice is the same as what is regularly used for the number of live points when using nested sampling in obtaining gravitational wave posterior~\citep{Smith:2019ucc}.

\subsection{Stopping Criteria}
\label{subsec:stop}
When multiplied by the likelihood value, the sampling weight gives the posterior distribution.
\begin{equation}
    \mathfrak{w}_k = w_k\,\mathcal{L}_k,
\end{equation}
the effective posterior-sample size is given by\footnote{Equal weighted samples can be obtained by performing rejection sampling using the normalised weights $\mathfrak{w}_k/\text{max}(\mathfrak{w}_k)$.}
\begin{equation}
n^\mathfrak{w}_\mathrm{eff} = \frac{\left(\sum_k \mathfrak{w}_k\right)^2}{\sum_k \left(\mathfrak{w}_k\right)^2}.
\end{equation}
Evidence normalises the posterior probability distribution and is thus just the sum of these weights,
\begin{equation}
    \mathcal{Z} = \sum_k \mathfrak{w}_k.
\end{equation}
Finally, the total probability enclosed by a likelihood threshold is,
\begin{align}
    \int_{\mathcal{L}(\boldsymbol{\theta}) > \mathcal{L}_*} \mathcal{L}(d | \boldsymbol{\theta}) \,
    \pi(\boldsymbol{\theta})\;\D\theta &\approx \nonumber \\
    \bar{P}(\mathcal{L}_*) =&\frac{\sum_k\mathfrak{w}_k\left(\mathcal{L}_k > \mathcal{L}_*\right)}{\sum_k\mathfrak{w}_k}.
    \label{eq:nested_prob}   
\end{align}
The numerator is summed only for samples with a likelihood value greater than $\mathcal{L}_*$.
Stopping criteria can be defined in several ways. Some examples when an analysis can be terminated include
\begin{enumerate}
    \item a chosen number of effective samples have been collected,
    \item increase in marginal likelihood stagnates,
    \item the probability enclosed by a likelihood threshold becomes smaller than a prechosen value.
\end{enumerate}
We stop the analysis for the presented analysis when $\bar{P} < 0.05$, i.e., the likelihood threshold encloses less than 5\% probability.

\subsection{Increase in Dimensionality}

\begin{center}
\begin{figure}
\includegraphics[width=0.49\textwidth]{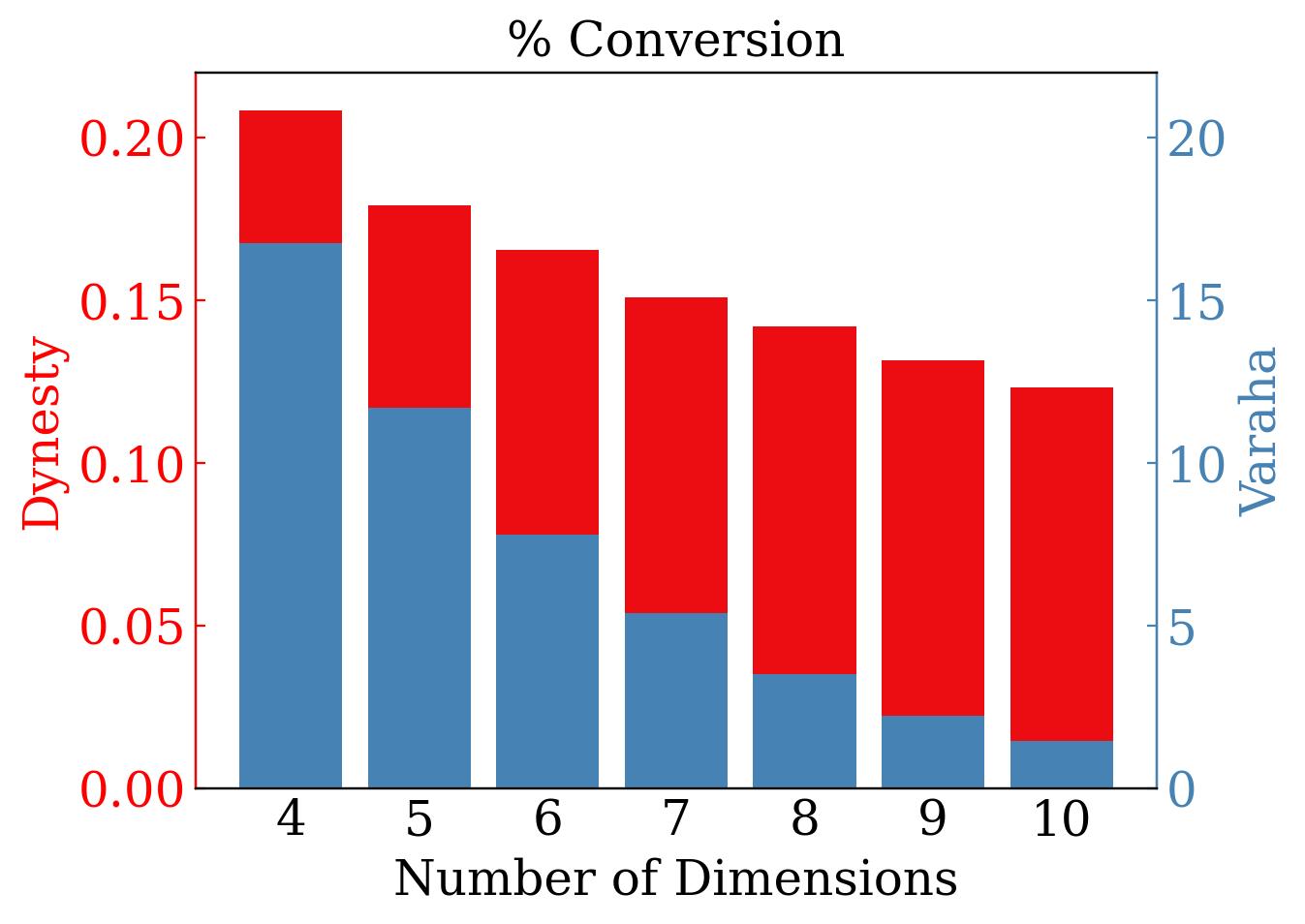}
\caption{Percentage conversion of points at which likelihood was calculated to effective posterior sample size. The sampled distribution is Rosenbrock. The sampling uses the condition described in Section~\ref{sec:results}.}
\label{fig:factors}
\end{figure}
\end{center}

Varaha becomes increasingly inefficient with the increase in dimensionality. For a $D$ dimensional space, if each dimension has the same number of bins, the number of bins in each dimension is $\sim \left( \delta \bar{V}(\mathcal{L}_*^i)\right)^{(-1/\text{D})}$. The number of bins becomes smaller for larger dimensionality, resulting in the live volume becoming significantly bigger than the true volume. The efficiency gained from storing all the points is gradually lost due to inefficient sampling of the parameter space. Fig.~\ref{fig:factors} shows the conversion percentage of the number of likelihood calculations to the posterior's effective sample size for Varaha and Dynesty~\citep{Speagle:2019ivv}. The chosen distribution is Rosenbrock~\citep{10.1093/comjnl/3.3.175, 10.1063/1.4819989}, a challenging sample distribution and often used in optimisation problems. The efficiency gradually decreases with increased dimensionality.

\begin{center}
\begin{figure}
\includegraphics[width=0.49\textwidth]{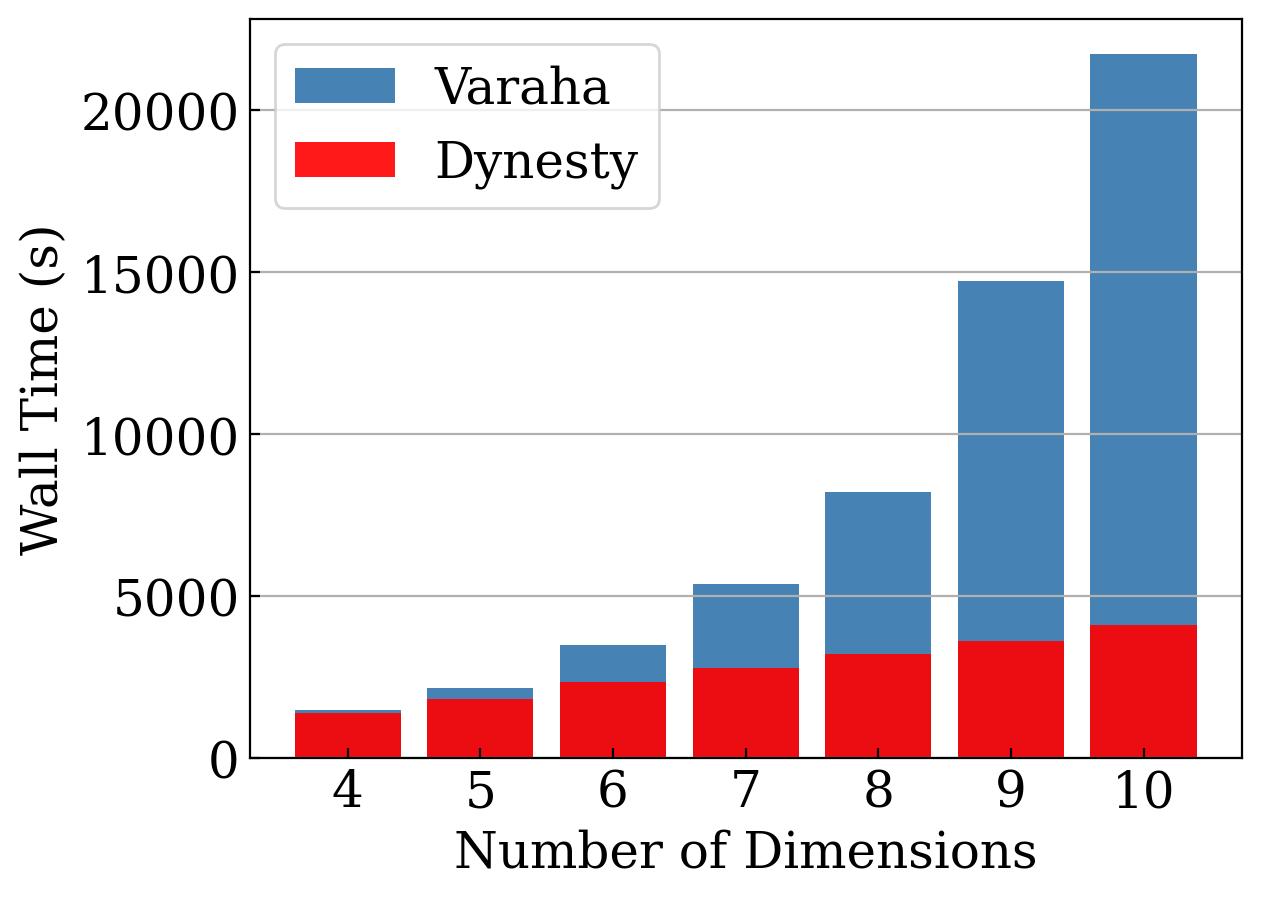}
\caption{Sampling wall time for Varaha compared to Dynesty using one CPU. The sampled distribution is Rosenbrock. The sampling uses the condition described in Section~\ref{sec:results}. Likelihood calculations for the Rosenbrock function contribute only a small fraction of the total computation.}
\label{fig:bookkeep}
\end{figure}
\end{center}

Increasing dimensionality also increases the number of cycle analyses take to complete. This results in an increase in computation needed to calculate the sampling density for each sample. Fig.~\ref{fig:bookkeep} shows the wall time for sampling the Rosenbrock distribution for different dimensionality. Each likelihood calculation takes a fraction of a millisecond, thus constituting a small fraction of the wall time. 
Varaha takes less time than Dynesty to collect one posterior sample for the dimensionalities discussed above. However, in the context of sampling \ac{GW} posteriors, the number of parameters for a precessing binary is 15. At this dimension, Varaha is inefficient and takes hundreds of millions of likelihood calculations to complete the sampling. Thus, its current implementation is not a sampler of choice. However, it is possible to separately sample extrinsic and intrinsic parameters and break one large dimensional problem into two small dimensional ones~\citep{2015PhRvD..92b3002P, 2018arXiv180510457L, 2023PhRvD.108b3001T}. Any dimensionality reduction improves Varaha's efficiency over Dynesty. Moreover, as the likelihood calculation is significantly more expensive, the wall time is dominated by likelihood calculations. 

\subsection{Parallelisation}
\begin{center}
\begin{figure}
\includegraphics[width=0.49\textwidth]{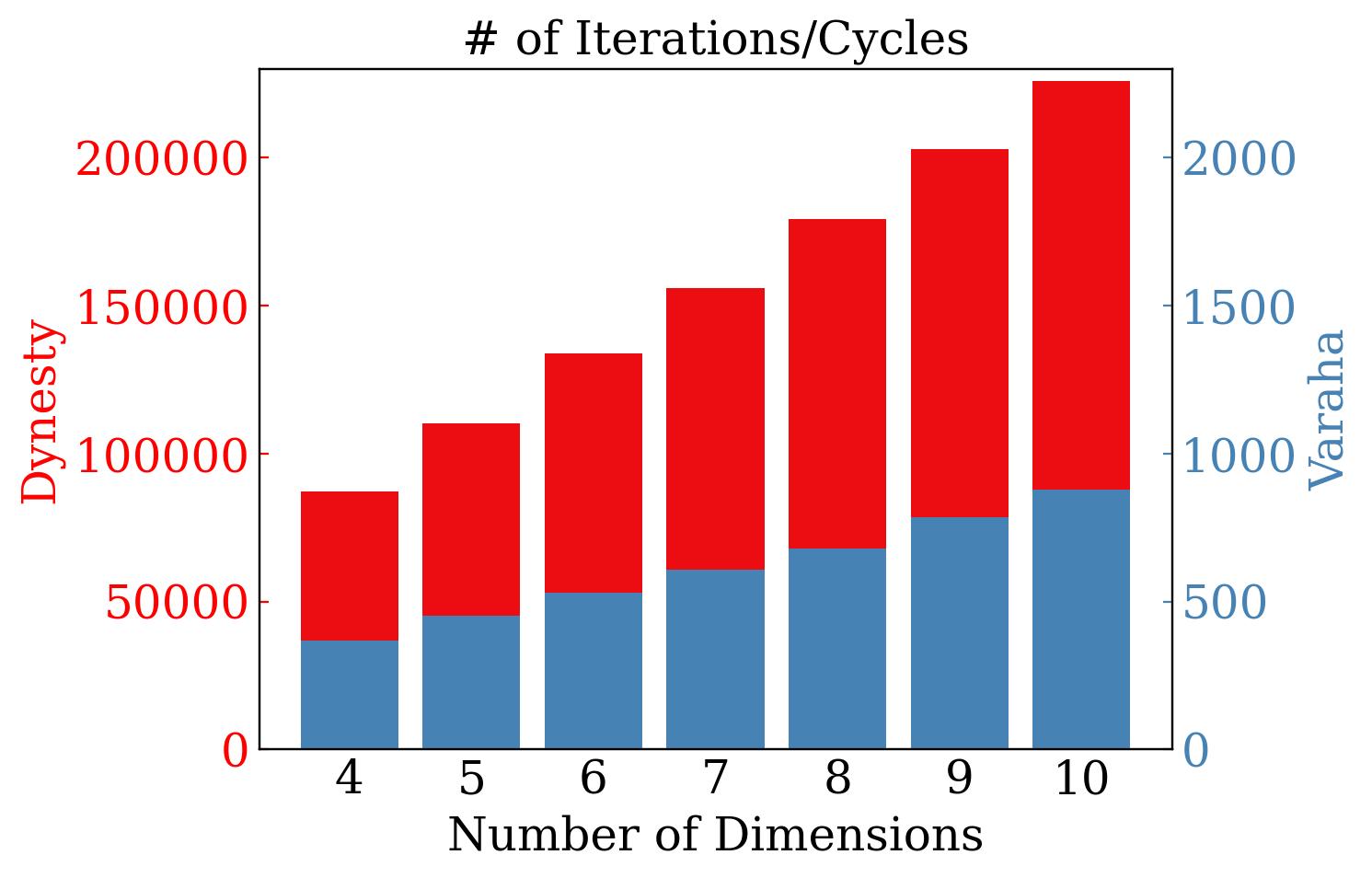}
\caption{Number of iterations/cycles made by Dynesty compared to  Varaha. The sampled distribution is Rosenbrock. The sampling uses the condition described at the end of subsection~\ref{subsec:stop} and sets $N_w=10,000$ and $n_w=2,000$.}
\label{fig:fparallel}
\end{figure}
\end{center}

An advantage of Varaha is that it efficiently parallels likelihood calculations over multiple processes. This is because of the significantly reduced likelihood cross-references in calculating the likelihood threshold compared to nested sampling. Fig.~\ref{fig:fparallel} shows the number of iterations/cycles used by Dynesty compared to Varaha. Varaha uses significantly less number of cycles. Each cycle calculates likelihood values over thousands of points, which can be parallelised over several CPUs. The calculation of sampling density is also parallelisable. Different nodes can evaluate sampling density for different bins, which can then be collected and added. 

\section{Sampling Examples}
\label{sec:results}

This section discusses a few examples to showcase Varaha's ability to sample challenging distributions. We also make comparisons with Nested Sampling. We have chosen seven dimensions for all the examples, explicitly aiming for the relevant dimensionality when \ac{GW} posteriors are obtained by separate sampling of extrinsic and intrinsic parameters.

Varaha shrank the volume by 95\% in each cycle for the examples discussed. The analysis proceeded to the next likelihood threshold only when the current likelihood threshold enclosed 2,000 reduced prior samples and 10,000 effective prior samples. We also sampled the distributions using Dynesty operated through the {\sc{bilby}} infrastructure~\citep{2019ApJS..241...27A}. We used the default settings provided by {\sc{bilby}} with some modifications to ensure reasonable convergence. Both samplers used a single CPU.

\subsection{Bi-Modal Multivariate}

We concoct a 7D bimodal distribution by combining two multivariate normal distributions. For the first normal, we randomly chose means between -2 and -1. The covariance was drawn from the Wishart distribution. The scale was set to a diagonal matrix with elements randomly chosen between 1. and 1.5. The elements were further divided by 56. For the second normal, we randomly chose means between 2 and 1. The covariance was drawn from the Wishart distribution. The scale was set to a diagonal matrix with elements randomly chosen between 1. and 1.5. The elements were further divided by 28. The mixing fraction was randomly chosen between 0.3 and 1.

\begin{center}
\begin{figure}
\includegraphics[width=0.49\textwidth]{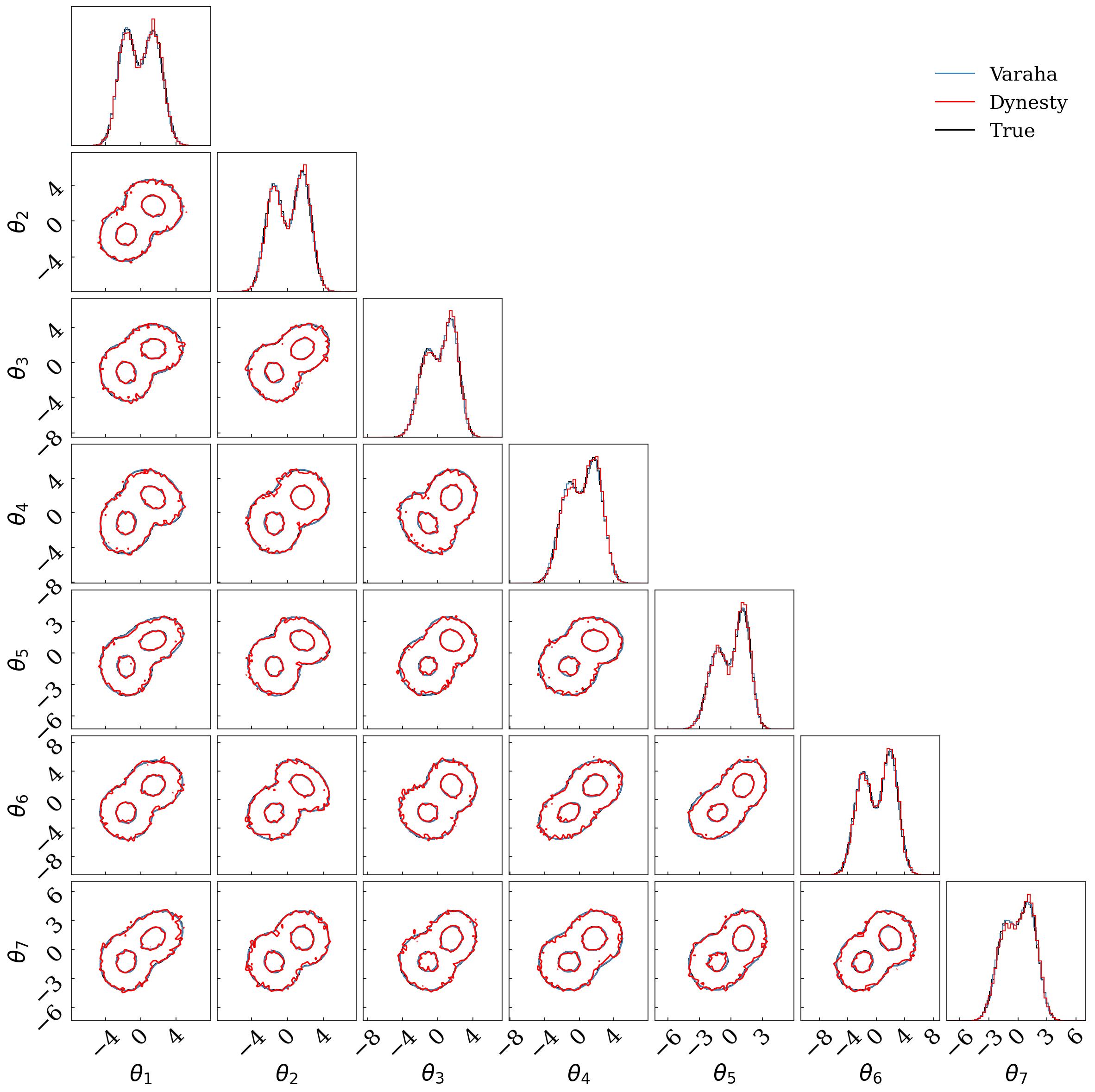}
\caption{Percentage conversion of points at which likelihood was calculated to effective posterior-sample size. The sampled distribution is Bi-modal Gaussian}
\label{fig:bimod}
\end{figure}
\end{center}

For both Dynesty and Varaha, we used a uniform prior between -20 and 20 for all the parameters. We used 2,000 live points for Dynesty. Dynesty collected 14,000 effective samples and estimated the log evidence to be $-25.824 \pm  0.091$. Varaha took 471 cycles. Varaha produced 6,41,300 effective samples and estimated log evidence to be $-25.823 \pm 0.015$. Evidence~(marginal likelihood) was calculated using bootstrapping. Varaha was around forty times more efficient in converting points where likelihood was calculated to posterior samples. Varaha took double the wall time compared to Dynesty. Likelihood calculation was cheap, thus most of the time was consumed by overhead computation. The comparison for the sampled distribution is shown in Fig.~\ref{fig:bimod}.

\subsection{Rosenbrock Distribution}
\label{subsec:rosen}

Rosenbrock is a challenging distribution to sample because of the high correlation between the parameters~\citep{Dittmann2024NotesOT}. 

\begin{center}
\begin{figure}
\includegraphics[width=0.49\textwidth]{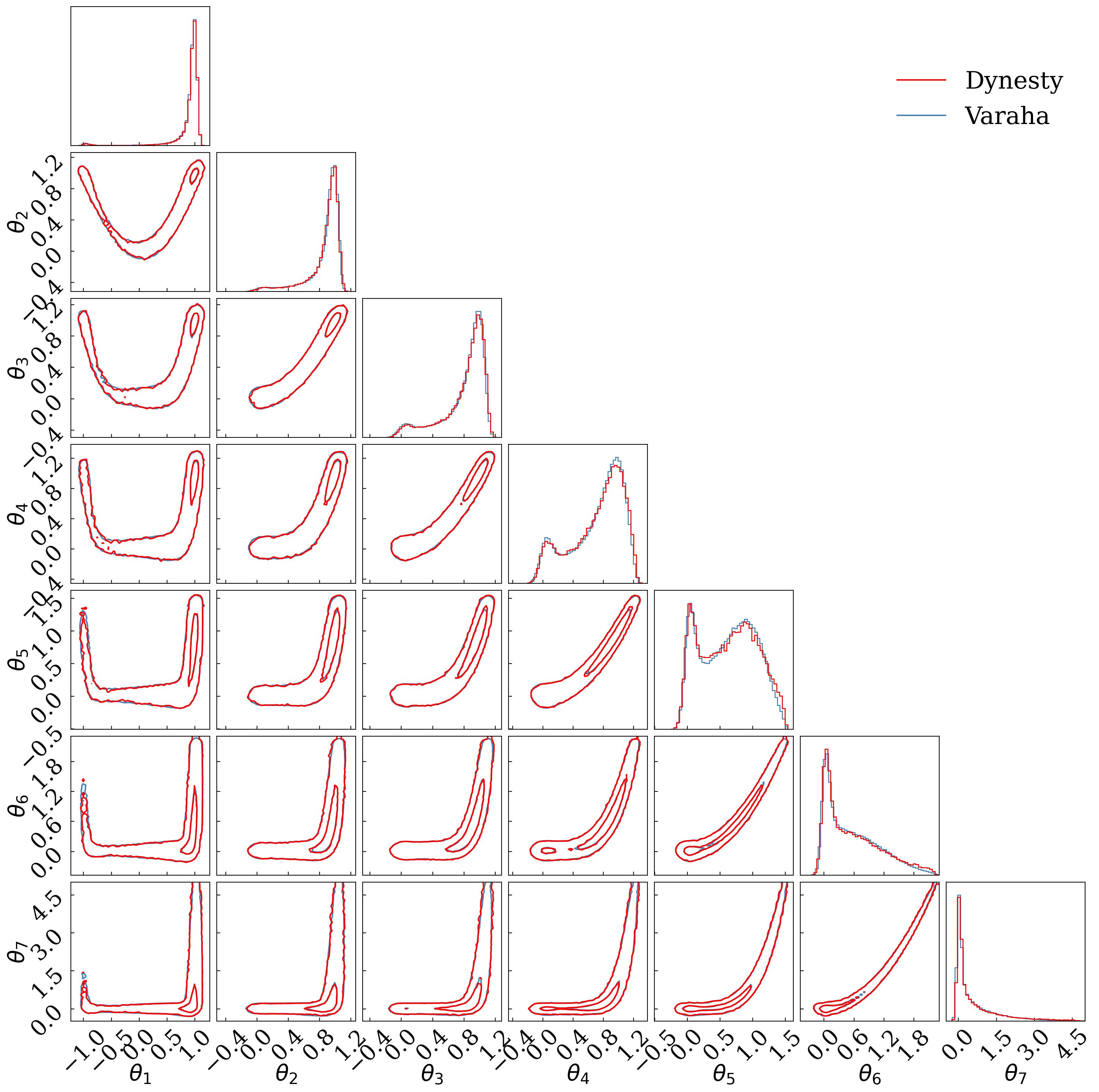}
\caption{Percentage conversion of points at which likelihood was calculated to effective posterior-sample size. The sampled distribution is Rosenbrock. The sampling uses the condition described in Section~\ref{sec:results}.}
\label{fig:rosen}
\end{figure}
\end{center}

For both Dynesty and Varaha, we used a uniform prior between -5 and 5 for all the parameters. Dynesty took 156,000 iterations. to complete. We used 5,000 live points. Dynesty collected 41,000 effective posterior-samples and estimated the log evidence to be $28.993 \pm  0.073$. Varaha took 586 cycles. Varaha collected 316,000 effective posterior samples and estimated log evidence to be $-29.05 \pm 0.0267$. Evidence~(marginal likelihood) was calculated using bootstrapping. Varaha was around forty times more efficient in converting points where likelihood was calculated to posterior samples. Varaha took double the wall time compared to Dynesty. Likelihood calculation was cheap; thus, most of the time, it was consumed by overhead computation. The comparison for the sampled distribution is shown in Fig.~\ref{fig:rosen}.

Our previous sampler version is significantly less efficient—the percentage conversion of the points where likelihood was calculated to effective posterior-sample size is only 0.1\% 

\subsection{Extrinsic Parameters for GW150914}

We sampled the extrinsic parameters listed in Table~\ref{tab:gw_params} for the first \ac{GW} signal GW150914~\citep{first_monday_pe}. We used an aligned spin model {\sc{IMRPhenomD}}~\citep{2016PhRvD..93d4006H, phenomd2}. This model ignores the components of the spins in the orbital plane of the binary, leaving only the component masses and the component of spins aligned with orbital angular momentum as the intrinsic parameters. For the component masses, we fixed, $m_1=36.80M_\odot$ and $m_2=31.96$, and the aligned spin components are fixed to $s_{1z} = -0.623$, $s_{2z} = 0.466$. We picked these values from a full sampling run separately performed using Dynesty. These choices correspond to the maximum likelihood value for that run. 

\begin{center}
\begin{figure}
\includegraphics[width=0.49\textwidth]{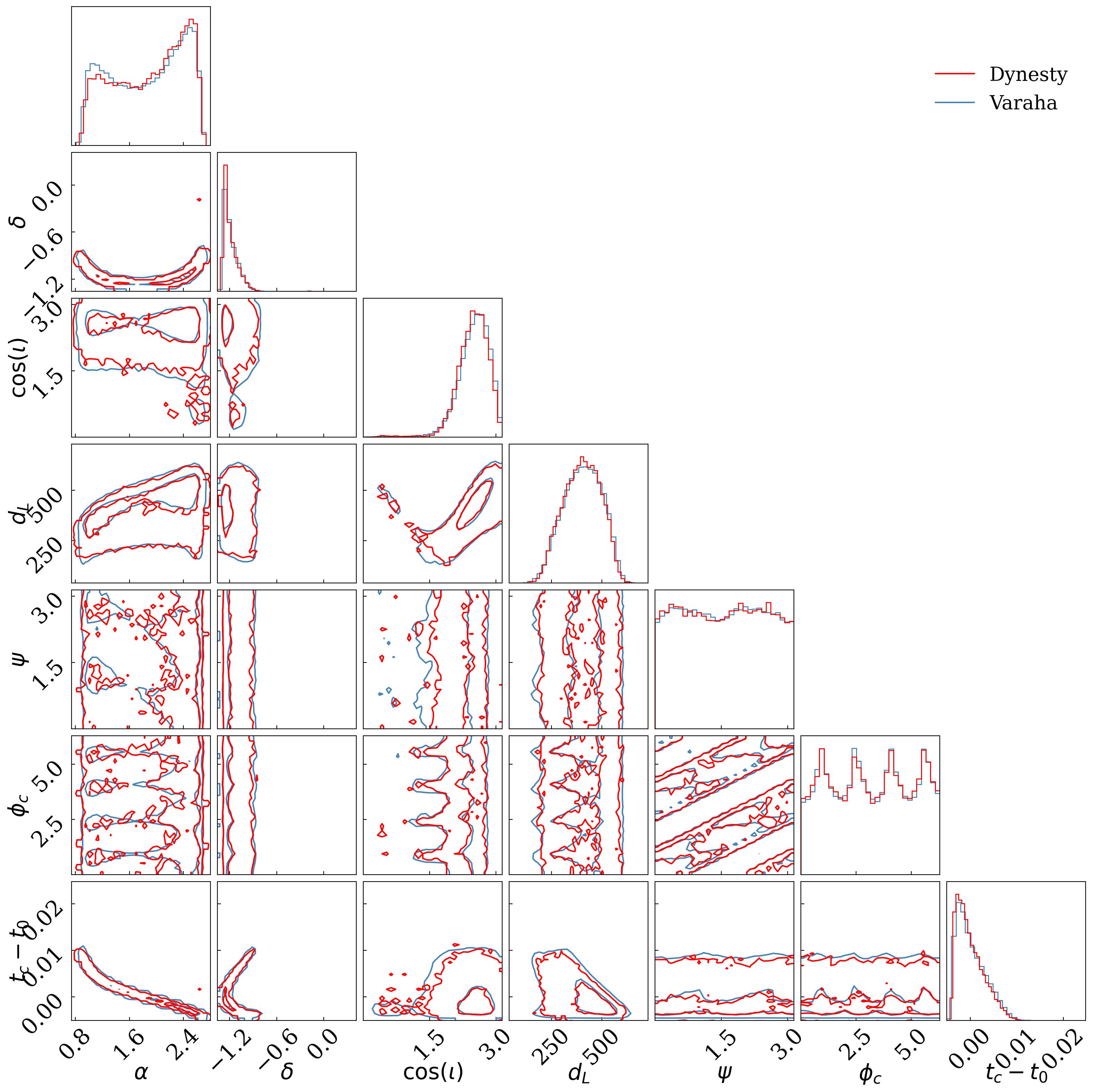}
\caption{The sampled extrinsic parameters detailed in Table~\ref{tab:gw_params} for the observations GW150914~\citep{first_monday_pe}. The mean coalescence time is $t_0=1126259462.411$.}
\label{fig:gw150914}
\end{figure}
\end{center}

We used 2,000 live points for Dynesty. Dynesty took 51,300 iterations to complete and collected 13,000 effective samples. Varaha took 1270 cycles to complete and collected 320,000 effective samples. Varaha was around twenty times more efficient in converting points where likelihood was calculated to posterior samples. Fig.\ref{fig:gw150914} shows the sampled extrinsic parameter.

\begin{table}
\caption{List of extrinsic parameters and the priors used estimated for the observation GW150914~\citep{first_monday_pe}. The GPS time $t_0=1126259462.411$.}
\begin{tabular}{| c | c | l |}
\hline
Parameter & Uniform Prior & Description \\
\hline
$\alpha$ & [0, 2$\pi$] & Right ascension of the source \\
$\sin \delta$ & [-1, 1]& Sine of declination of the source \\
$d_L$ & [0, 5$\times10^3$]Mpc & Luminosity distance of the source \\
$\cos \iota$ & [-1, 1] & Cosine of the inclination angle \\
$\psi$ & [0, $\pi$] & Polarisation angle \\
$\phi_{c}$ & [0, 2$\pi$] & Coalescence phase \\
$t_c$ & $t_0$ + [0.1, -0.1] & Coalescence time in the reference detector \\
\hline
\end{tabular} 
\label{tab:gw_params}
\end{table}

\subsection{Intrinsic Parameters for GW190412}
\label{subsec:gw190412}
 The parameter estimation is performed using a Bayesian setup. The posterior on the intrinsic parameters, $\theta$ and extrinsic parameters, $\Omega$ is obtained using the Bayes equation,
\begin{equation}
    p\left(\theta, \Omega\,|\,\vec d\right) = \frac{\mathcal{L}\left(\vec d\,|\,\theta, \Omega\right)\,p\left(\theta, \Omega\right)}{p\left(\vec d \right)},
    \label{eq:bayespe1}
\end{equation}
where $p(\theta, \Omega)$ is the prior on the parameters, $\mathcal{L}\left(\vec d\,|\,\theta, \Omega\right)$ is the likelihood and $p(\vec d)$ is the normalisation constant. Equation~\ref{eq:bayespe1} is estimated by sampling the posterior which removes the need to calculate the normalisation. Moreover, this equation can be integrated over the extrinsic parameters to reduce the sampling to the intrinsic parameters,
\begin{equation}
    p\left(\theta|\,\vec d\right) \propto \mathcal{L}_\mathrm{red}\left(\vec d\,|\,\theta, \right)\,p\left(\theta, \Omega\right),\, \mathcal{L}_\mathrm{red} = \int_\Omega \mathcal{L}\left(\vec d \,|\, \Omega, \theta\right)\, \mathrm{d}\Omega.
    \label{eq:bayespe2}
\end{equation}
As various modes mix differently at different inclination angles, this variable is included with the intrinsic parameters when sampling. The total number of parameters sampled is, therefore, 9, and the reduced likelihood $\mathcal{L}_\mathrm{red}$ for a fixed value of these parameters is obtained by integrating over the remaining six parameters. This marginalisation can be done in many ways. A Monte Carlo scheme would sample the extrinsic parameters, $\Omega_{ij}$ for a fixed set of intrinsic parameters, $\theta_i$ and sum the corresponding likelihood values to obtain reduced likelihood,
\begin{equation}
    \mathcal{L}_\mathrm{red}\left(\theta_i\,|\,\vec d\right) = \sum_j \mathcal{L}\left(\Omega_{ij}\,|\, \vec \theta_i, d\right).
    \label{eq:redlik}
\end{equation}
Samples $\Omega_{ij}$, correspond to the posterior $p(\Omega\,|\,\theta_i,\,\vec d)$. However, as extrinsic parameters have a weak dependence on the intrinsic parameters, instead of integrating likelihood over the full extrinsic parameter space, for the examples presented here, we only integrate over parts of the extrinsic parameter space that meaningfully contribute to the posterior. We identify this space by sampling the extrinsic parameters for a fiducial choice of intrinsic parameters and recalculate the likelihood for different values of $\theta_i$. As we are not repeatedly sampling the extrinsic parameter space for different choices of $\theta_i$, although approximate, this approach is computationally cheap. 

\begin{center}
\begin{figure}
\includegraphics[width=0.49\textwidth]{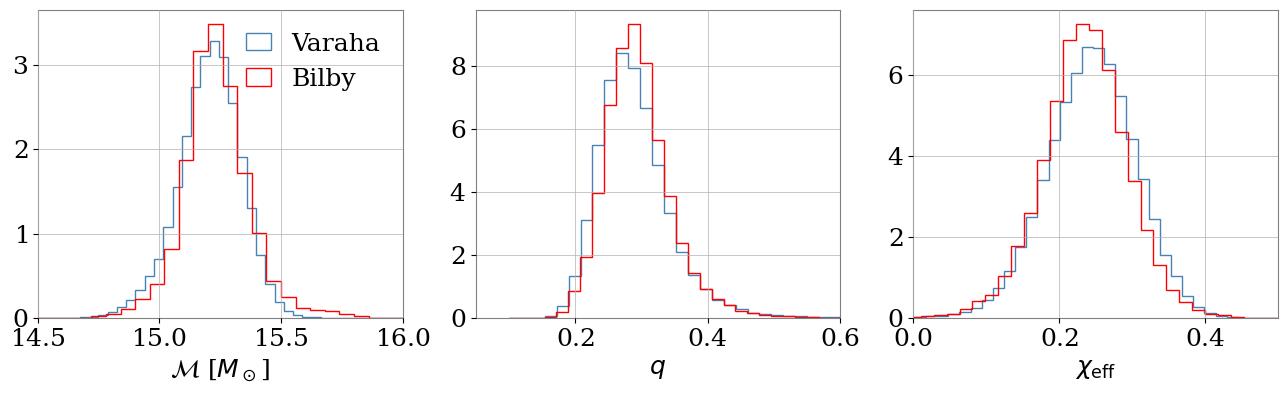}
\caption{The sampled intrinsic parameters detailed in Table~\ref{tab:gw_params_intirn} for the observation GW190412~\citep{LIGOScientific:2020stg}. This figure also shows the posterior obtained using Dynesty operated through the {\sc{bilby}} infrastructure. However, we sampled over the 13 dimensions~(15 dimensions marginalised over distance and coalescence time.). The posterior from two samplers looks consistent. Varaha sampled on a 9-dimensional space using reduced likelihoods and was two orders of magnitude more efficient, but the computational cost was added when marginalising over extrinsic parameters.}
\label{fig:gw190412}
\end{figure}
\end{center}

We sample the intrinsic parameters listed in Table~\ref{tab:gw_params_intirn} for the \ac{GW} signal GW190412~\citep{LIGOScientific:2020stg}. The components for this \ac{BBH} were asymmetric in masses and were moderately spinning. We used the waveform model {\sc{IMRPhenomXPHM}} that incorporates spin precession and higher harmonics~\citep{2021PhRvD.103j4056P}. We first sample the extrinsic parameters. For the component masses, we just fix, $m_1=34.78M_\odot$ and $m_2=9.08$, and the aligned spin components to $s_{1z} = 0.32$, $s_{2z} = -0.133$. We picked these values from a full sampling run separately performed using Dynesty. These choices correspond to the maximum likelihood value for that run. Sampling of intrinsic parameters is performed using the reduced likelihood obtained by marginalising over the extrinsic parameters. A relevant space for the extrinsic parameters is generated by retaining the luminosity distance, right ascension, and declination and combining them with samples drawn from the full range of polarisation angle, coalescence phase, and coalescence time. The samples in the extrinsic space have different likelihood values corresponding to each sample of the intrinsic parameters. The reduced likelihood is just the sum of all these values~\citep{2023PhRvD.108b3001T}. The volume enclosing most of the posterior probability mass in the $\alpha\text{--}\delta\text{--}d_L$ space changes with the change in the intrinsic parameters. However, this change is not significant; intrinsic parameters of interest have similar signal morphology. Therefore, the marginalisation is done on a volume that is double the volume that encloses 99.9\% posterior probability mass. Apparently, this volume is represented by the samples collected in the first step when performing sampling over the extrinsic parameters. The number of samples impacts the error on the reduced likelihood calculation. Using a large sample size will reduce statistical errors. For the presented analysis, we generated an extrinsic space that resulted in statistical errors between 1--2\%. Figure~\ref{fig:gw190412} shows the posterior on the chirp mass, mass ratio and the effective spin~($\chi_\mathrm{eff} = (s_{1z} + q\,s_{2z}) / (1 + q)$). We do not show the posterior on the extrinsic parameters as these need to be reconstructed once the posterior on the intrinsic parameters have been obtained~\citep{2020PASA...37...36T}. We have not verified if the choices are robust for a large class of signals, and more work is needed to investigate these issues carefully. 

\begin{table}
\caption{List of intrinsic parameters and the priors used for the observation GW190412~\citep{LIGOScientific:2020stg}. Along with the masses and spins of a compact binary, the angles on the cone of precession about the total
angular momentum completely defines the system.}
\begin{tabular}{| c | c | l |}
\hline
Parameter & Uniform Prior & Description\\
\hline
$\mathcal{M}$ & [12.4, 17.3] & Chirp Mass \\
$q$ & [0.1, 1.0] & Mass Ratio\\
$a_{1,2}$ & [-1, 1] & Spin magnitude of the two object\\
$\cos \theta_{1, 2}$ & [0, 1.0] & Tilt angle\\
$\phi_{12}$ & [0, 2$\pi$] & Separation between spin's azimuthal angles\\
$\phi_{JL}$ & [0, 2$\pi$] & Azimuth of orbital angular momentum\\
\hline
\end{tabular} 
\label{tab:gw_params_intirn}
\end{table}

\subsection{Population Level Test on Intrinsic Parameters}
\label{subsec:pop}

We assess the accuracy and efficiency of the sampler by performing a population-level test on intrinsic parameters. For a choice of detector frame chirp mass~(the redshifted chirp mass value measured by the \ac{GW} data), aligned spin and mass ratio values, we draw extrinsic parameters as described in \ref{tab:gw_params}. The chosen parameters are used to create synthetic injections. In this analysis, the network is comprised of advanced LIGO Livingston and Hanford detectors \cite{2015CQGra..32g4001L}. Any injection that crosses a matched filter network \ac{SNR} of 10 is identified as \emph{observed} and selected for estimating the parameters. The intrinsic parameter draws ensure that the \emph{observed} distribution, which is given in Table~\ref{tab:intr_param}, matches the prior. Varaha first estimates the extrinsic parameters by fixing the intrinsic parameters to the drawn value. Once extrinsic parameters have been estimated, intrinsic parameters are estimated by sampling on reduced likelihood; likelihood value marginalised over the extrinsic parameter samples~\citep{2023PhRvD.108b3001T}.

\begin{table}
\caption{List of intrinsic parameters and the priors used for performing the population test. The detector frame chirp mass is \emph{observed} chirp mass value caused due to redshifting by the expanding universe. The injections that cross a network \ac{SNR} of 10.0 are tagged as \emph{observed}. The prior and the \emph{observed} parameters are uniformly distributed. The two aligned spin components are drawn independently.}
\begin{tabular}{| l | l | l |}
\hline
Parameter & Uniform Prior & Description\\
\hline
$\mathcal{M}$ & [5, 50]$M_\odot$ & Chirp Mass\\
$q$ & [0.1, 1.0] & Mass Ratio\\
$s_{1z}$ & [-1, 1] & Aligned Spin for the first component\\
$s_{2z}$ & [-1, 1] & Aligned Spin for the second component\\
\hline
\end{tabular} 
\label{tab:intr_param}
\end{table}

\begin{figure}[t!]
    \centering
    \includegraphics[width=0.45\textwidth]{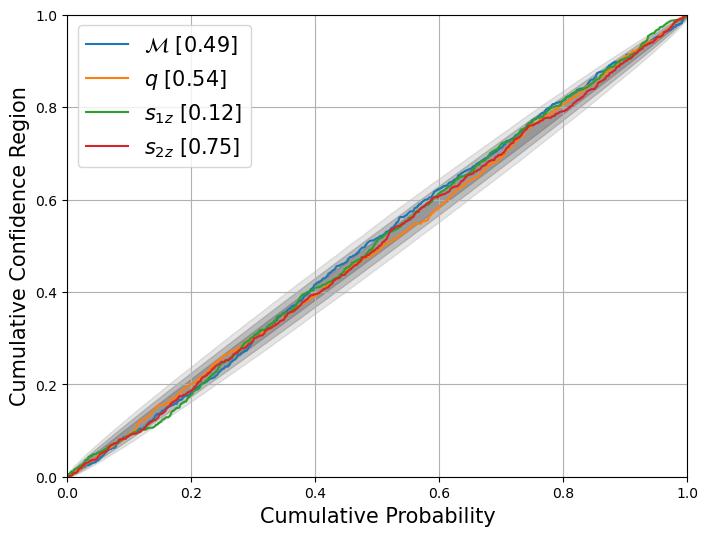}
    \caption{Percentile-Percentile (P-P) plot for 500 simulated injections. The grey bands indicate the 1-, 2- and 3-$\sigma$ confidence intervals. For results to be unbiased, the trails are required to be enclosed by the bands~\citep{2014PhRvD..89h4060S}. The brackets enclose the p-value in each case.}
    \label{fig:pp}
\end{figure}

The P-P test investigates if the measured interval of parameters at a credibility $f\%$ also encloses $f\%$ of true values among all the measurements. This test identifies any population-level biases in the measurement of parameters. We perform parameter estimation on 1,000 injections. The P-P plot for the intrinsic parameters shown in Figure~\ref{fig:pp} presents no noticeable bias.

The average efficiency for the parameter estimation runs was around 6\%. The average number of samples collected by the runs was approximately 200,000. This example can be repeated after including in-plane spins, thus increasing the sampling space to 8 dimensions. Referring back to Figure~\ref{fig:factors}, we expect Varaha to be more than an order of magnitude efficient than nested sampling in this case.



\subsection{Comparison With the Previous Version}

\begin{center}
\begin{figure}
\includegraphics[width=0.49\textwidth]{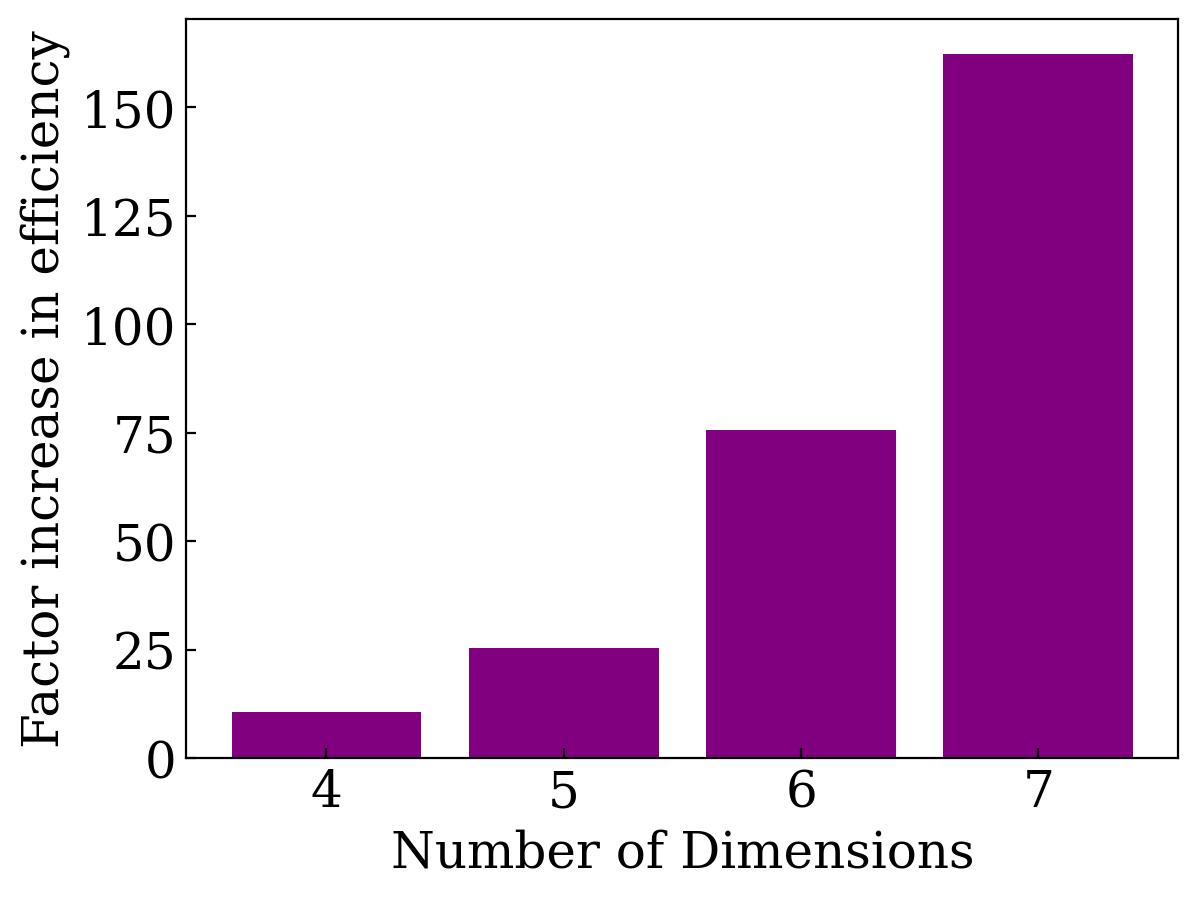}
\caption{Factor increase in efficiency between the presented sampler and the previous version of Varaha. Efficiency is defined as the ratio of posterior sample size and the number of likelihood calculations. The presented version of Varaha is significantly more efficient.}
\label{fig:v1vsv2}
\end{figure}
\end{center}

Compared to the previous version of Varaha~\citep{2023PhRvD.108b3001T}, the presented version is significantly more efficient. The increase in efficiency is due to sampling the regions with higher likelihood in the parameter space. In Figure~\ref{fig:v1vsv2} we show an increase in efficiency when sampling the Rosenbrock function. We get a comparable increase in efficiency in different dimensions for the bi-modal example. For sampling extrinsic parameters of GW150914 the efficiency increases by around two orders of magnitude. The presented sampler introduces an added cost to calculate the sampling density. Therefore, the old version is a sampler of choice when sampling a distribution in small dimensions with cheap likelihood calculation. This is applicable when sampling is performed to estimate extrinsic parameters for a compact binary after fixing the intrinsic parameters~\citep{2015PhRvD..92b3002P}. However, for expensive likelihoods, the presented likelihood is better suited which is applicable when sampling the intrinsic parameters using the reduced likelihood. For the previous case, a likelihood calculation usually takes a fraction of a millisecond, while for the latter, it is expensive by three orders of magnitude or more. For the latter case, the time taken to calculate sampling density constitutes a small fraction of the total sampling time. Therefore, a pragmatic setup is to use the old version of Varaha to estimate the reduced likelihood and the presented version to sample the intrinsic parameters using the reduced likelihood.

\section{Conclusions}

This article presented a novel sampler significantly more efficient than Nested sampling for small dimensions~($\leq$ 10). We built on previous work and showed that it is possible to explicitly calculate the sampling density of all the sampled points in small dimensions. This sampling density, along with the likelihood values, constitutes the posterior distribution. Our sampler, Varaha, does not discard any points, making it more efficient than nested sampling, but as it becomes increasingly inefficient with increased dimensionality, it is not a sampler of choice to sample the full parameter space of compact binaries. However, if the sampling is separated into observer-dependent parameters and parameters intrinsic to the binary, we expect Varaha to be a sampler of choice. It is efficient and embarrassingly parallel, and as it retains all the samples, posterior corresponding to different priors can be efficiently obtained by re-weighting the sampling density. The presented version of Varaha is significantly more efficient than the previous version~\citep{2023PhRvD.108b3001T}, better suited for cheaper likelihoods~(e.g. when marginalising over extrinsic parameters). The new version comes with an added cost of estimating the sampling density, but this cost constitutes only a small fraction of the total computational cost for more expensive likelihoods~(e.g. when sampling intrinsic parameters using marginalised likelihoods, which can take from a fraction of a second to a few seconds). The presented sampler is lightweight, requiring less than four hundred lines of code.

Varaha has multiple avenues for improvement. We did not thoroughly investigate the most optimum setting to maximise the efficiency. We also did not investigate an optimum methodology for creating multi-dimensional bins. We verified that the method gives unbiased estimates for aligned-spin binaries; however, this method needs to be verified for a class of signals.  

\section*{Acknowledgements}

Thanks to Lalit Pathak for LIGO's Publications \& Presentations review, Alberto Vecchio, Patricia Schmidt and Stephen Fairhurst for helpful guidance on this project, and Charlie Hoy for providing Varaha's interface for Bilby. This work was conducted on Cardiff University's HAWK HPC.

This material is based upon work supported by NSF's LIGO Laboratory which is a major facility fully funded by the National Science Foundation.

\section*{Data Availability}

The examples and required code are available on \href{https://github.com/vaibhavtewari/Varaha}{GitHub}.



\bibliographystyle{mnras}
\bibliography{references} 

\begin{thebibliography}{}
\makeatletter
\relax
\def\mn@urlcharsother{\let\do\@makeother \do\$\do\&\do\#\do\^\do\_\do\%\do\~}
\def\mn@doi{\begingroup\mn@urlcharsother \@ifnextchar [ {\mn@doi@} {\mn@doi@[]}}
\def\mn@doi@[#1]#2{\def\@tempa{#1}\ifx\@tempa\@empty \href {http://dx.doi.org/#2} {doi:#2}\else \href {http://dx.doi.org/#2} {#1}\fi \endgroup}
\def\mn@eprint#1#2{\mn@eprint@#1:#2::\@nil}
\def\mn@eprint@arXiv#1{\href {http://arxiv.org/abs/#1} {{\tt arXiv:#1}}}
\def\mn@eprint@dblp#1{\href {http://dblp.uni-trier.de/rec/bibtex/#1.xml} {dblp:#1}}
\def\mn@eprint@#1:#2:#3:#4\@nil{\def\@tempa {#1}\def\@tempb {#2}\def\@tempc {#3}\ifx \@tempc \@empty \let \@tempc \@tempb \let \@tempb \@tempa \fi \ifx \@tempb \@empty \def\@tempb {arXiv}\fi \@ifundefined {mn@eprint@\@tempb}{\@tempb:\@tempc}{\expandafter \expandafter \csname mn@eprint@\@tempb\endcsname \expandafter{\@tempc}}}

\bibitem[\protect\citeauthoryear{{Aasi}, {Abbott}, {Abbott}, {Abbott}  et~al.}{{Aasi} et~al.}{2015}]{2015CQGra..32g4001L}
{Aasi} J.,  {Abbott} B.~P.,  {Abbott} R.,  {Abbott} T.,   et~al., 2015, \mn@doi [Classical and Quantum Gravity] {10.1088/0264-9381/32/7/074001}, \href {https://ui.adsabs.harvard.edu/\\#abs/2015CQGra..32g4001L} {32, 074001}

\bibitem[\protect\citeauthoryear{Abbott et~al.}{Abbott et~al.}{2016}]{first_monday_pe}
Abbott B.~P.,  et~al., 2016, Phys. Rev. Lett., 116, 241102

\bibitem[\protect\citeauthoryear{{Abbott} et~al.}{{Abbott} et~al.}{2020a}]{2020LRR....23....3A}
{Abbott} R.,  et~al., 2020a, \mn@doi [Living Reviews in Relativity] {10.1007/s41114-020-00026-9}, \href {https://ui.adsabs.harvard.edu/abs/2020LRR....23....3A} {23, 3}

\bibitem[\protect\citeauthoryear{Abbott et~al.}{Abbott et~al.}{2020b}]{LIGOScientific:2020stg}
Abbott R.,  et~al., 2020b, \mn@doi [Phys. Rev. D] {10.1103/PhysRevD.102.043015}, 102, 043015

\bibitem[\protect\citeauthoryear{{Abbott}, {Abbott}, {Acernese}  et~al.}{{Abbott} et~al.}{2021a}]{2021arXiv210801045T}
{Abbott} R.,  {Abbott} T.~D.,  {Acernese} F.,   et~al., 2021a, arXiv e-prints, \href {https://ui.adsabs.harvard.edu/abs/2021arXiv210801045T} {p. arXiv:2108.01045}

\bibitem[\protect\citeauthoryear{{Abbott}, {Abbott}, {Acernese}  et~al.}{{Abbott} et~al.}{2021b}]{o3b_cat}
{Abbott} R.,  {Abbott} T.~D.,  {Acernese} F.,   et~al., 2021b, arXiv e-prints, \href {https://arxiv.org/abs/2111.03606} {p. arXiv:2111.03606}

\bibitem[\protect\citeauthoryear{{Adamcewicz}, {Lasky}  \& {Thrane}}{{Adamcewicz} et~al.}{2023}]{2023ApJ...958...13A}
{Adamcewicz} C.,  {Lasky} P.~D.,   {Thrane} E.,  2023, \mn@doi [\apj] {10.3847/1538-4357/acf763}, \href {https://ui.adsabs.harvard.edu/abs/2023ApJ...958...13A} {958, 13}

\bibitem[\protect\citeauthoryear{{Ashton} et~al.,}{{Ashton} et~al.}{2019}]{2019ApJS..241...27A}
{Ashton} G.,  et~al., 2019, \mn@doi [The Astrophysical Journal Supplement Series] {10.3847/1538-4365/ab06fc}, \href {https://ui.adsabs.harvard.edu/abs/2019ApJS..241...27A} {241, 27}

\bibitem[\protect\citeauthoryear{{Callister} \& {Farr}}{{Callister} \& {Farr}}{2024}]{2024PhRvX..14b1005C}
{Callister} T.~A.,  {Farr} W.~M.,  2024, \mn@doi [Physical Review X] {10.1103/PhysRevX.14.021005}, \href {https://ui.adsabs.harvard.edu/abs/2024PhRvX..14b1005C} {14, 021005}

\bibitem[\protect\citeauthoryear{{Dax}, {Green}, {Gair}, {Macke}, {Buonanno}  \& {Sch{\"o}lkopf}}{{Dax} et~al.}{2021}]{2021PhRvL.127x1103D}
{Dax} M.,  {Green} S.~R.,  {Gair} J.,  {Macke} J.~H.,  {Buonanno} A.,   {Sch{\"o}lkopf} B.,  2021, \mn@doi [\prl] {10.1103/PhysRevLett.127.241103}, \href {https://ui.adsabs.harvard.edu/abs/2021PhRvL.127x1103D} {127, 241103}

\bibitem[\protect\citeauthoryear{Dittmann}{Dittmann}{2024}]{Dittmann2024NotesOT}
Dittmann A.~J.,  2024. \url {https://api.semanticscholar.org/CorpusID:269430717}

\bibitem[\protect\citeauthoryear{{Fairhurst}, {Hoy}, {Green}, {Mills}  \& {Usman}}{{Fairhurst} et~al.}{2023}]{2023PhRvD.108h2006F}
{Fairhurst} S.,  {Hoy} C.,  {Green} R.,  {Mills} C.,   {Usman} S.~A.,  2023, \mn@doi [\prd] {10.1103/PhysRevD.108.082006}, \href {https://ui.adsabs.harvard.edu/abs/2023PhRvD.108h2006F} {108, 082006}

\bibitem[\protect\citeauthoryear{{Farah}, {Edelman}, {Zevin}, {Fishbach}, {Mar{\'\i}a Ezquiaga}, {Farr}  \& {Holz}}{{Farah} et~al.}{2023}]{2023ApJ...955..107F}
{Farah} A.~M.,  {Edelman} B.,  {Zevin} M.,  {Fishbach} M.,  {Mar{\'\i}a Ezquiaga} J.,  {Farr} B.,   {Holz} D.~E.,  2023, \mn@doi [\apj] {10.3847/1538-4357/aced02}, \href {https://ui.adsabs.harvard.edu/abs/2023ApJ...955..107F} {955, 107}

\bibitem[\protect\citeauthoryear{Feroz \& Skilling}{Feroz \& Skilling}{2013}]{10.1063/1.4819989}
Feroz F.,  Skilling J.,  2013, \mn@doi [AIP Conference Proceedings] {10.1063/1.4819989}, 1553, 106

\bibitem[\protect\citeauthoryear{{Gabbard}, {Messenger}, {Heng}, {Tonolini}  \& {Murray-Smith}}{{Gabbard} et~al.}{2022}]{2022NatPh..18..112G}
{Gabbard} H.,  {Messenger} C.,  {Heng} I.~S.,  {Tonolini} F.,   {Murray-Smith} R.,  2022, \mn@doi [Nature Physics] {10.1038/s41567-021-01425-7}, \href {https://ui.adsabs.harvard.edu/abs/2022NatPh..18..112G} {18, 112}

\bibitem[\protect\citeauthoryear{{Gupta}}{{Gupta}}{2024}]{2024arXiv240207075G}
{Gupta} I.,  2024, \mn@doi [arXiv e-prints] {10.48550/arXiv.2402.07075}, \href {https://ui.adsabs.harvard.edu/abs/2024arXiv240207075G} {p. arXiv:2402.07075}

\bibitem[\protect\citeauthoryear{{Heinzel}, {Biscoveanu}  \& {Vitale}}{{Heinzel} et~al.}{2023}]{2023arXiv231200993H}
{Heinzel} J.,  {Biscoveanu} S.,   {Vitale} S.,  2023, \mn@doi [arXiv e-prints] {10.48550/arXiv.2312.00993}, \href {https://ui.adsabs.harvard.edu/abs/2023arXiv231200993H} {p. arXiv:2312.00993}

\bibitem[\protect\citeauthoryear{{Husa}, {Khan}, {Hannam}, {P{\"u}rrer}, {Ohme}, {Forteza}  \& {Boh{\'e}}}{{Husa} et~al.}{2016}]{2016PhRvD..93d4006H}
{Husa} S.,  {Khan} S.,  {Hannam} M.,  {P{\"u}rrer} M.,  {Ohme} F.,  {Forteza} X.~J.,   {Boh{\'e}} A.,  2016, \mn@doi [\prd] {10.1103/PhysRevD.93.044006}, \href {https://ui.adsabs.harvard.edu/\\#abs/2016PhRvD..93d4006H} {93, 044006}

\bibitem[\protect\citeauthoryear{Khan et~al.}{Khan et~al.}{2016}]{phenomd2}
Khan S.,  et~al., 2016, Phys. Rev. D, 93

\bibitem[\protect\citeauthoryear{{Lange}, {O'Shaughnessy}  \& {Rizzo}}{{Lange} et~al.}{2018}]{2018arXiv180510457L}
{Lange} J.,  {O'Shaughnessy} R.,   {Rizzo} M.,  2018, arXiv e-prints, \href {https://ui.adsabs.harvard.edu/abs/2018arXiv180510457L} {p. arXiv:1805.10457}

\bibitem[\protect\citeauthoryear{{Leyde}, {Green}, {Toubiana}  \& {Gair}}{{Leyde} et~al.}{2024}]{2024PhRvD.109f4056L}
{Leyde} K.,  {Green} S.~R.,  {Toubiana} A.,   {Gair} J.,  2024, \mn@doi [\prd] {10.1103/PhysRevD.109.064056}, \href {https://ui.adsabs.harvard.edu/abs/2024PhRvD.109f4056L} {109, 064056}

\bibitem[\protect\citeauthoryear{{Maga{\~n}a Hernandez} \& {Ray}}{{Maga{\~n}a Hernandez} \& {Ray}}{2024}]{2024arXiv240402522M}
{Maga{\~n}a Hernandez} I.,  {Ray} A.,  2024, \mn@doi [arXiv e-prints] {10.48550/arXiv.2404.02522}, \href {https://ui.adsabs.harvard.edu/abs/2024arXiv240402522M} {p. arXiv:2404.02522}

\bibitem[\protect\citeauthoryear{Martino, Victor  \& Carlos}{Martino et~al.}{2017}]{ESS}
Martino L.,  Victor E.,   Carlos S.,  2017, \mn@doi [Signal Processing-Elsevier] {https://doi.org/10.1016/j.sigpro.2016.08.025}, pp 386--401

\bibitem[\protect\citeauthoryear{{Morisaki}}{{Morisaki}}{2021}]{2021PhRvD.104d4062M}
{Morisaki} S.,  2021, \mn@doi [\prd] {10.1103/PhysRevD.104.044062}, \href {https://ui.adsabs.harvard.edu/abs/2021PhRvD.104d4062M} {104, 044062}

\bibitem[\protect\citeauthoryear{{Morisaki} \& {Raymond}}{{Morisaki} \& {Raymond}}{2020}]{2020PhRvD.102j4020M}
{Morisaki} S.,  {Raymond} V.,  2020, \mn@doi [\prd] {10.1103/PhysRevD.102.104020}, \href {https://ui.adsabs.harvard.edu/abs/2020PhRvD.102j4020M} {102, 104020}

\bibitem[\protect\citeauthoryear{{Morisaki}, {Smith}, {Tsukada}, {Sachdev}, {Stevenson}, {Talbot}  \& {Zimmerman}}{{Morisaki} et~al.}{2023}]{2023PhRvD.108l3040M}
{Morisaki} S.,  {Smith} R.,  {Tsukada} L.,  {Sachdev} S.,  {Stevenson} S.,  {Talbot} C.,   {Zimmerman} A.,  2023, \mn@doi [\prd] {10.1103/PhysRevD.108.123040}, \href {https://ui.adsabs.harvard.edu/abs/2023PhRvD.108l3040M} {108, 123040}

\bibitem[\protect\citeauthoryear{{Morr{\'a}s}, {Nu{\~n}o Siles}  \& {Garc{\'\i}a-Bellido}}{{Morr{\'a}s} et~al.}{2023}]{2023PhRvD.108l3025M}
{Morr{\'a}s} G.,  {Nu{\~n}o Siles} J.~F.,   {Garc{\'\i}a-Bellido} J.,  2023, \mn@doi [\prd] {10.1103/PhysRevD.108.123025}, \href {https://ui.adsabs.harvard.edu/abs/2023PhRvD.108l3025M} {108, 123025}

\bibitem[\protect\citeauthoryear{{Narola}, {Janquart}, {Meijer}, {Haris}  \& {Van Den Broeck}}{{Narola} et~al.}{2023}]{2023arXiv230812140N}
{Narola} H.,  {Janquart} J.,  {Meijer} Q.,  {Haris} K.,   {Van Den Broeck} C.,  2023, \mn@doi [arXiv e-prints] {10.48550/arXiv.2308.12140}, \href {https://ui.adsabs.harvard.edu/abs/2023arXiv230812140N} {p. arXiv:2308.12140}

\bibitem[\protect\citeauthoryear{{Pankow}, {Brady}, {Ochsner}  \& {O'Shaughnessy}}{{Pankow} et~al.}{2015}]{2015PhRvD..92b3002P}
{Pankow} C.,  {Brady} P.,  {Ochsner} E.,   {O'Shaughnessy} R.,  2015, \mn@doi [\prd] {10.1103/PhysRevD.92.023002}, \href {https://ui.adsabs.harvard.edu/abs/2015PhRvD..92b3002P} {92, 023002}

\bibitem[\protect\citeauthoryear{{Pathak}, {Reza}  \& {Sengupta}}{{Pathak} et~al.}{2023}]{2023PhRvD.108f4055P}
{Pathak} L.,  {Reza} A.,   {Sengupta} A.~S.,  2023, \mn@doi [\prd] {10.1103/PhysRevD.108.064055}, \href {https://ui.adsabs.harvard.edu/abs/2023PhRvD.108f4055P} {108, 064055}

\bibitem[\protect\citeauthoryear{{Pathak}, {Munishwar}, {Reza}  \& {Sengupta}}{{Pathak} et~al.}{2024}]{2024PhRvD.109b4053P}
{Pathak} L.,  {Munishwar} S.,  {Reza} A.,   {Sengupta} A.~S.,  2024, \mn@doi [\prd] {10.1103/PhysRevD.109.024053}, \href {https://ui.adsabs.harvard.edu/abs/2024PhRvD.109b4053P} {109, 024053}

\bibitem[\protect\citeauthoryear{{Payne}, {Talbot}, {Lasky}, {Thrane}  \& {Kissel}}{{Payne} et~al.}{2020}]{2020PhRvD.102l2004P}
{Payne} E.,  {Talbot} C.,  {Lasky} P.~D.,  {Thrane} E.,   {Kissel} J.~S.,  2020, \mn@doi [\prd] {10.1103/PhysRevD.102.122004}, \href {https://ui.adsabs.harvard.edu/abs/2020PhRvD.102l2004P} {102, 122004}

\bibitem[\protect\citeauthoryear{{Payne}, {Kremer}  \& {Zevin}}{{Payne} et~al.}{2024}]{2024ApJ...966L..16P}
{Payne} E.,  {Kremer} K.,   {Zevin} M.,  2024, \mn@doi [\apjl] {10.3847/2041-8213/ad3e82}, \href {https://ui.adsabs.harvard.edu/abs/2024ApJ...966L..16P} {966, L16}

\bibitem[\protect\citeauthoryear{{Pratten} et~al.,}{{Pratten} et~al.}{2021}]{2021PhRvD.103j4056P}
{Pratten} G.,  et~al., 2021, \mn@doi [\prd] {10.1103/PhysRevD.103.104056}, \href {https://ui.adsabs.harvard.edu/abs/2021PhRvD.103j4056P} {103, 104056}

\bibitem[\protect\citeauthoryear{{Ray}, {Hernandez}, {Mohite}, {Creighton}  \& {Kapadia}}{{Ray} et~al.}{2023}]{2023ApJ...957...37R}
{Ray} A.,  {Hernandez} I.~M.,  {Mohite} S.,  {Creighton} J.,   {Kapadia} S.,  2023, \mn@doi [\apj] {10.3847/1538-4357/acf452}, \href {https://ui.adsabs.harvard.edu/abs/2023ApJ...957...37R} {957, 37}

\bibitem[\protect\citeauthoryear{{Rinaldi}, {Del Pozzo}, {Mapelli}, {Lorenzo-Medina}  \& {Dent}}{{Rinaldi} et~al.}{2024}]{2024A&A...684A.204R}
{Rinaldi} S.,  {Del Pozzo} W.,  {Mapelli} M.,  {Lorenzo-Medina} A.,   {Dent} T.,  2024, \mn@doi [\aap] {10.1051/0004-6361/202348161}, \href {https://ui.adsabs.harvard.edu/abs/2024A&A...684A.204R} {684, A204}

\bibitem[\protect\citeauthoryear{Rosenbrock}{Rosenbrock}{1960}]{10.1093/comjnl/3.3.175}
Rosenbrock H.~H.,  1960, \mn@doi [The Computer Journal] {10.1093/comjnl/3.3.175}, 3, 175

\bibitem[\protect\citeauthoryear{{Roulet}, {Mushkin}, {Wadekar}, {Venumadhav}, {Zackay}  \& {Zaldarriaga}}{{Roulet} et~al.}{2024}]{2024arXiv240402435R}
{Roulet} J.,  {Mushkin} J.,  {Wadekar} D.,  {Venumadhav} T.,  {Zackay} B.,   {Zaldarriaga} M.,  2024, \mn@doi [arXiv e-prints] {10.48550/arXiv.2404.02435}, \href {https://ui.adsabs.harvard.edu/abs/2024arXiv240402435R} {p. arXiv:2404.02435}

\bibitem[\protect\citeauthoryear{{Sadiq}, {Dent}  \& {Gieles}}{{Sadiq} et~al.}{2024}]{2024ApJ...960...65S}
{Sadiq} J.,  {Dent} T.,   {Gieles} M.,  2024, \mn@doi [\apj] {10.3847/1538-4357/ad0ce6}, \href {https://ui.adsabs.harvard.edu/abs/2024ApJ...960...65S} {960, 65}

\bibitem[\protect\citeauthoryear{{Sidery}, {Aylott}, {Christensen}  et~al.}{{Sidery} et~al.}{2014}]{2014PhRvD..89h4060S}
{Sidery} T.,  {Aylott} B.,  {Christensen} N.,   et~al., 2014, \mn@doi [\prd] {10.1103/PhysRevD.89.084060}, \href {https://ui.adsabs.harvard.edu/abs/2014PhRvD..89h4060S} {89, 084060}

\bibitem[\protect\citeauthoryear{{Smith}, {Ashton}, {Vajpeyi}  \& {Talbot}}{{Smith} et~al.}{2020a}]{2020MNRAS.498.4492S}
{Smith} R. J.~E.,  {Ashton} G.,  {Vajpeyi} A.,   {Talbot} C.,  2020a, \mn@doi [\mnras] {10.1093/mnras/staa2483}, \href {https://ui.adsabs.harvard.edu/abs/2020MNRAS.498.4492S} {498, 4492}

\bibitem[\protect\citeauthoryear{Smith, Ashton, Vajpeyi  \& Talbot}{Smith et~al.}{2020b}]{Smith:2019ucc}
Smith R. J.~E.,  Ashton G.,  Vajpeyi A.,   Talbot C.,  2020b, \mn@doi [Mon. Not. Roy. Astron. Soc.] {10.1093/mnras/staa2483}, 498, 4492

\bibitem[\protect\citeauthoryear{Speagle}{Speagle}{2020}]{Speagle:2019ivv}
Speagle J.~S.,  2020, \mn@doi [Mon. Not. Roy. Astron. Soc.] {10.1093/mnras/staa278}, 493, 3132

\bibitem[\protect\citeauthoryear{{The LIGO-Virgo-KAGRA (LVK) Collaboration}}{{The LIGO-Virgo-KAGRA (LVK) Collaboration}}{2023}]{2023arXiv230408393T}
{The LIGO-Virgo-KAGRA (LVK) Collaboration} 2023, \mn@doi [arXiv e-prints] {10.48550/arXiv.2304.08393}, \href {https://ui.adsabs.harvard.edu/abs/2023arXiv230408393T} {p. arXiv:2304.08393}

\bibitem[\protect\citeauthoryear{{Thrane} \& {Talbot}}{{Thrane} \& {Talbot}}{2020}]{2020PASA...37...36T}
{Thrane} E.,  {Talbot} C.,  2020, \mn@doi [\pasa] {10.1017/pasa.2020.23}, \href {https://ui.adsabs.harvard.edu/abs/2020PASA...37...36T} {37, e036}

\bibitem[\protect\citeauthoryear{{Tiwari}}{{Tiwari}}{2018}]{2018CQGra..35n5009T}
{Tiwari} V.,  2018, \mn@doi [Classical and Quantum Gravity] {10.1088/1361-6382/aac89d}, \href {https://ui.adsabs.harvard.edu/\\#abs/2018CQGra..35n5009T} {35, 145009}

\bibitem[\protect\citeauthoryear{{Tiwari}}{{Tiwari}}{2024}]{2024MNRAS.527..298T}
{Tiwari} V.,  2024, \mn@doi [\mnras] {10.1093/mnras/stad3155}, \href {https://ui.adsabs.harvard.edu/abs/2024MNRAS.527..298T} {527, 298}

\bibitem[\protect\citeauthoryear{{Tiwari}, {Hoy}, {Fairhurst}  \& {MacLeod}}{{Tiwari} et~al.}{2023}]{2023PhRvD.108b3001T}
{Tiwari} V.,  {Hoy} C.,  {Fairhurst} S.,   {MacLeod} D.,  2023, \mn@doi [\prd] {10.1103/PhysRevD.108.023001}, \href {https://ui.adsabs.harvard.edu/abs/2023PhRvD.108b3001T} {108, 023001}

\bibitem[\protect\citeauthoryear{{Williams}, {Veitch}  \& {Messenger}}{{Williams} et~al.}{2021}]{2021PhRvD.103j3006W}
{Williams} M.~J.,  {Veitch} J.,   {Messenger} C.,  2021, \mn@doi [\prd] {10.1103/PhysRevD.103.103006}, \href {https://ui.adsabs.harvard.edu/abs/2021PhRvD.103j3006W} {103, 103006}

\bibitem[\protect\citeauthoryear{{Wong}, {Isi}  \& {Edwards}}{{Wong} et~al.}{2023}]{2023ApJ...958..129W}
{Wong} K. W.~K.,  {Isi} M.,   {Edwards} T. D.~P.,  2023, \mn@doi [\apj] {10.3847/1538-4357/acf5cd}, \href {https://ui.adsabs.harvard.edu/abs/2023ApJ...958..129W} {958, 129}

\makeatother
\end{thebibliography}




\appendix

\section{Filling the Bins}

The choice of bin size and the number of samples drawn from bins substantially impact the evolution of the analysis. The number of bins in each dimension results in each hyper-cube occupying a volume in the parameter space. The requirement is to have this volume close to the error in volume. We did not make significant progress in optimising these elements. The analysis draws the number of bins in each dimension using the following ad hoc prescription,
\begin{equation}
B^i_j = B\,2^{U(-1, 1)}\quad\mathrm{or}\quad\mathcal{N}(\mu=B, \sigma=\sqrt{B}),\quad i=\mathrm{fixed}, j = 1\cdots D 
\label{nbins}
\end{equation}
where $U(-1, 1)$ is a uniform draw between -1 and 1, $\mathcal{N}$ is the normal distribution and $B = \left(\delta \bar{V}(\mathcal{L}_*^i)\right)^{(-1/\text{D})}$. One draw of $D$ bins is not necessarily expected to have a hypercube volume close to $\delta \bar{V}(\mathcal{L}_*^i)$. Thus, we make several thousands of draws and choose the combination that has hypercube volume closest to $\delta \bar{V}(\mathcal{L}_*^i)$. Moreover, we switch between two prescriptions separated by the or condition in Eq.~\ref{nbins} depending on whether the cycle number is odd or even.

At a cycle, the number of samples needs to be increased such that shrink criteria defined in~\ref{sec:method} is met. The parameter space is binned, and bins that contain at least one sample with a likelihood threshold greater than $\mathcal{L}_*^i$ are selected. The number of samples in bins is adjusted such that the summed sampling density in the bins is approximately uniform. The number of samples to be drawn from each bin is
\begin{equation}
    N_j^i \propto B_j^i - \max(B_j^i),\quad B_j^i = \sum_k \rho_k(\mathcal{L}_k > \mathcal{L}^i_*)\,\Theta(j, k),
\end{equation}
where the summed density in a bin, $B_j^i$ is the sum of the sampling density of all the samples that cross the likelihood threshold $\mathcal{L}_*^i$ and overlap with the $j^{th}$ bin. 

Each cycle approximately shrinks the volume by a factor $f$. This approximately implies that the number of samples enclosed by the new likelihood threshold is also $f$ times the number of samples enclosed by the old likelihood threshold. Thus, the total number of samples that need to be compensated is $1-f$ times the total number of samples. As not all samples drawn from bins will cross the likelihood threshold, this needs to be divided by the acceptance fraction~(ratio of samples that cross the likelihood threshold and the total number of samples drawn from the bins). Starting at the first cycle, where almost all the points sampled from the parameter space cross the likelihood threshold, the acceptance fraction reduces as likelihood increases.

\section{Sampling density calculation}

The sampling density calculation requires adding contributions from all the bins sampled in all the cycles. This has to be done for each sample. Each cycle shrinks the volume and proceeds to sample regions with higher likelihood in the parameter space. Thus, removing the non-contributing samples and bins from earlier cycles speeds up the calculation of the sampling density. The analysis keeps all the samples enclosed by a volume twice the volume of the current cycle. For the current cycle $i$, any samples with a likelihood value smaller than $L^\mathrm{keep}_*$ are removed. $L^\mathrm{keep}_*$ is chosen, such that
\begin{equation}
    \frac{\bar{V}^\mathrm{keep}}{\bar{V}^i} > 2.
\end{equation}
With increase in $\mathcal{L}^i_*$, there is increase in $\mathcal{L}_*^\mathrm{keep}$. To avoid removing samples meaningfully contributing to the posterior, the analysis does not increase $L^\mathrm{keep}_*$ if the following condition is met
\begin{equation}
    \frac{\sum_k\mathfrak{w}_k\left(\mathcal{L}_k > \mathcal{L}^i_*\right)}{\sum_k\mathfrak{w}_k\left(\mathcal{L}_k > \mathcal{L}_*^\mathrm{keep}\right)} < 0.999;
\end{equation}
the analysis stores all the samples enclosed by a volume that is double the volume that encloses 99.9\% posterior probability mass.

Varaha creates and samples from bins at each likelihood threshold, $\mathcal{L}_*^i$. Bins created at a volume 32 times bigger than $\bar{V}^\mathrm{keep}$ are removed. This also requires the removal of samples drawn from these bins~(these samples will have a likelihood greater than $\mathcal{L}^\mathrm{keep}$). Moreover, the contribution to the sampling density for any samples overlapping with the removed bins also needs to be subtracted. 

\bsp	
\label{lastpage}
\end{document}